\documentclass[seceq]{ptptex}

\usepackage{graphicx}
\usepackage{wrapft}

\title{
PCAC Relation and Pion Production-Absorption in Nuclei
}

\author{
Tetsuro \textsc{Nogi}, Toru \textsc{Sato} and Hisao \textsc{Ohtsubo}
}

\inst{
Department of Physics, Graduate School of Science, Osaka University,\\
Toyonaka, Osaka 560-0043, Japan
}

\abst{

Nuclear PCAC relation is studied in the framework of the effective theory of
nuclear interaction, in which the interaction of  real pion 
production-absorption is expressed by many-body operators, and does not 
include the one-nucleon operator as was assumed in the conventional works, while the effective
 axial-vector current includes  the one-nucleon  current in contrast to the former interaction. This problem is investigated  under the simple linear $\sigma$-model. Results are as folows: 1) The theory  describes consistently the PCAC relation and the pion production-absorption process. 2) The conventional  interpretation of the  effective pion source
 function as the interaction Hamiltonian of pion production-absorption does not hold. 3) The effective pion source function still includes the one-nucleon operator for the pion production-absorption at threshold effectively, which may justify the conventional theory.
}
\newcommand{\sbm}[1]{ \mbox{\boldmath{$#1$}} }

\begin{document}

\maketitle

\section{Introduction}
It is important to understand how the axial-vector current is modified in nuclei.
So far this modification has been described by introducing the meson-exchange current,
i.e,. many-body operators or renormalization of the coupling constants
in the current.(See the review article in Ref. (1).)
The former is convenient to be handled in finite nuclei, while the latter, in the nuclear
matter. As is well known, the axial-vector current does not conserve, and obeys the PCAC
 relation that divergence of the axial-vector current is proportional to the pion field.
The matrix element of the PCAC relation between nuclear states can be connected with
the matrix element of pion production-absorption in nuclei. The divergence
of the one-nucleon axial-vector current is
expressed by the p-wave pion-nucleon interaction. Therefore, it is
natural to assume that the two-nucleon axial-vector current is connected with the
pion production-absorption by two-nucleons.
 This idea was at first adopted for the
 exchange current of Gamow-Teller type by Blin-Stoyle and Tint \cite{Blyn67}, who assumed
the p-wave pion production operator of the one-nucleon type and also of the 
 two-nucleon type. Structure of the latter type is to be determined experimentally.
Since then this idea has been accepted  by many people. Among them   Bernab$\rm{\acute{e}}$u et al. mentioned\cite{Ber69} a similarity between the
muon capture reaction at large energy transfer
 and the pion production in nuclei at threshold both in kinematics and
 in dynamics, and investigated in detail the renormalization of the
 axial-vector current in nuclei.

Recently Kobayashi et al.\cite{Kobayashi97} pointed out that the one-nucleon operator 
 for pion production-absorption does not exist in the effective nuclear interaction
 Hamiltonian, when we introduce the one-pion-exchange potential into the nuclear
  effective Hamiltonian. They proposed  the theory of 
the effective interactions of
pions and nucleons in nuclei by adopting the unitary transformation which enables us
to describe the nuclear system with the energy above the pion production threshold. 
Similar work was done by Shebeko and Shirokov\cite{She00}.
It has been argued that the PCAC relation gives a consistency relation
among nuclear potential, axial-vector current and pion
production operator\cite{Dmi98}, and that
 the one-nucleon axial-vector current 
is connected with the interaction between a pion and a nucleon. 
One may say that absence of the one-nucleon operator for pion
 production-absorption  contradicts with the PCAC relation.
 
 The purpose of the present paper is to show that absence of the 
one-body operator does not contradict with the PCAC relation.
 We shall investigate the PCAC relations in nuclei in connection with the pion production-absorption process in two-body system within
 the effective theory by the unitary transformation.
 To make our discussion clear, we shall adopt the linear $\sigma$-model\cite{sigma}
 which satisfies the PCAC relation, and investigate the effective
 nuclear
 interactions within  the one-boson-exchange approximation,
 although the conclusion does not depend upon details of the model.

In \S 2, we shall derive the effective nuclear Hamiltonian in the
 framework of linear $\sigma$-model
under the unitary transformation. To make our story simple and clear,
 we restrict ourselves to the 
 effective interactions with up to the one-boson-exchange. Explicit
 forms of the nuclear force and the operators of pion
 production-absorption  will be derived. In \S 3, 
we shall transform the basic axial-vector current into the effective
nuclear axial-vector current
by adopting the same unitary transformation
  as employed in the previous section, and show
 that the effective
 nuclear axial-vector current
includes meson currents in addition to  the one-nucleon
 and the two-nucleon currents.
In \S 4 and \S 5, it will turn out that the effective nuclear axial-vector
 current thus derived satisfies the PCAC relation, and that the pion production by the two-nucleon operators in the present theory is consistent with the PCAC relation.
The results will be summarized in \S 6.

\section{ Effective Nuclear Interaction}

\subsection{Basic Model of the Strong Interaction}

We start from the Lagrangian of the linear $\sigma$-model,

\begin{eqnarray}
{L}&=&\int d\sbm{x}{\cal L}(\sbm{x})
\end{eqnarray}
with
\begin{eqnarray}
{\cal L}(\sbm{x})&=&\bar{\psi}(\sbm{x})[i\gamma_\mu \partial^\mu - M_N]\psi(\sbm{x}) \nonumber \\
&& +\frac{1}{2}[(\partial^\mu\sigma(\sbm{x}))^2 - m_\sigma^2 \sigma(\sbm{x})^2] + \frac{1}{2}[(\partial^\mu \vec{\pi}(\sbm{x}))^2 - m_\pi^2\vec{\pi}(\sbm{x})^2] \nonumber \\
&&-g\bar{\psi}(\sbm{x})[\sigma(\sbm{x}) + i\gamma_5\vec{\tau}\cdot\vec{\pi}(\sbm{x})]\psi(\sbm{x}) \nonumber \\
&&-\frac{m_\sigma^2-m_\pi^2}{2f_\pi}\sigma(\sbm{x})[\sigma(\sbm{x})^2 + \vec{\pi}(\sbm{x})^2]
-\frac{m_\sigma^2-m_\pi^2}{8f_\pi^2}[\sigma(\sbm{x})^2 + \vec{\pi}(\sbm{x})^2]^2,
\end{eqnarray}
where  the Goldberger-Treiman relation reads as $gf_\pi = M_N$ and $g_A = 1$.
Following the standard procedure we obtain the Hamiltonian $H$:

\begin{eqnarray}
H=H_0+H_I,
\label{eqn:sigmaM}
\end{eqnarray}
where $H_0$ is the free Hamiltonian of mesons and nucleons. The
basic interaction Hamiltonian $H_I$, which is relevant to the effective
nuclear interaction Hamiltonian in the one-boson-exchange approximation,
 consists of $\pi NN$, $\sigma NN$ and $\sigma
\pi\pi$ interactions as follows:
\begin{eqnarray}
H_I & = & H_{\pi NN}+H_{\sigma NN}+H_{\sigma\pi\pi}\\
H_{\pi NN} & = & ig \int d\sbm{x}
\bar{\psi}(\sbm{x})\gamma_5
\vec{\tau}\cdot\vec{\pi}(\sbm{x})\psi(\sbm{x}),\\ 
H_{\sigma NN}&=& g \int d\sbm{x}
     \bar{\psi}(\sbm{x}) \sigma(\sbm{x})\psi(\sbm{x}),\\ 
H_{\sigma\pi\pi} &=& \frac{m_\sigma^2-m_\pi^2}{2f_\pi}\int d\sbm{x}
        \sigma(\sbm{x})\vec{\pi}(\sbm{x})^2.
\end{eqnarray}

The relevant axial-vector current is expressed as 

\begin{eqnarray}
  \sbm{A}^a(\sbm{x})&=&
  \ \bar{\psi}(\sbm{x})\sbm{\gamma}\gamma_{5}\frac{\tau^a}2
  \psi(\sbm{ x})
  -\sigma(\sbm{ x})\sbm{\nabla}\pi^a(\sbm{x})
  +\pi^a(\sbm{x})  \sbm{\nabla}\sigma(\sbm{ x})
  -f_{\pi}\sbm{\nabla}\pi^a(\sbm{x}), \label{ax_s}\\
  A^{0a}(\sbm{x})&=&
  \ \bar{\psi}(\sbm{x})\gamma_0\gamma_{5}\frac{\tau^a}2  \psi(\sbm{ x})
  +\sigma(\sbm{ x})  \Pi_\pi^a(\sbm{x})
  -\pi^a(\sbm{x})  \Pi_\sigma(\sbm{ x})  +f_{\pi}\Pi_\pi^a(\sbm{x}),\label{ax_t}
\end{eqnarray}
where $\Pi_\pi^a$ and $\Pi_\sigma$ are conjugate momenta of $\pi$
 field and $\sigma$ field, respectively, and the superscript $a$ stands for
the isospin index.

\subsection{The Nuclear Force and Interaction of Pion Production-absorption}

\begin{wrapfigure}{r}{7cm}

\centerline{\includegraphics[width=7.5cm]{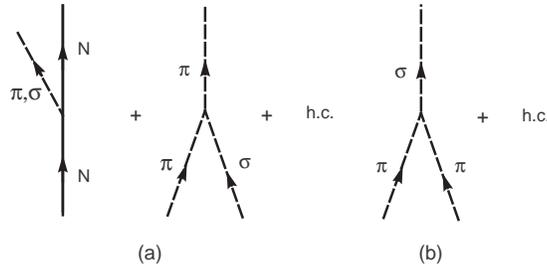}}
\caption{Decomposition of the interactions 
 into the interactions of "virtual" processes (a) and "real" processes (b).
The solid and broken lines stand for nucleons and mesons, respectively. }
\label{fig:1}
\end{wrapfigure}
We derive the effective nuclear Hamiltonian by using the
 the unitary transformation described in Ref. \cite{Kobayashi97}.
We define the nuclear Hamiltonian as follows:
\begin{eqnarray}
H_{eff} & = & U^\dagger H U.
\end{eqnarray}
Here, the unitary transformation 
$U=\exp(iS)\exp(iS')$ 
is chosen so as to eliminate the interactions, which correspond to the 'virtual'
processes, from the Hamiltonian, while the interactions which correspond to
 the 'real' processes are kept in the transformed Hamiltonian.
The interactions for 'virtual processes' of $H_I$  in our present model
are shown in Fig. 1(a), while 
those for the 'real processes'($H_{\sigma\pi\pi}^P$), in Fig. 1(b). 
Then, the Hamiltonian transformed by $\exp(iS)$ newly involves the interactions
 which correspond to higher order "virtual" processes. A typical term
  is shown in Fig. 2, which should be eliminated by the second 
 order unitary transformation  $\exp(iS')$.

 These observations fix the unitary transformation as follows:
\begin{eqnarray}
H_{\pi NN} + H_{\sigma NN} + H_{\sigma \pi \pi}^Q
 + [H_0, iS] & = & 0, \\
\frac{1}{2}[H_I + H_{\sigma\pi\pi}^P,iS]^Q + [H_0,iS'] = 0,
\end{eqnarray}
where the superscripts $P$ and $Q$ stand for the interactions
corresponding to 'real' and 'virtual' processes, respectively.
Then the effective Hamiltonian is given as
\begin{eqnarray}
H_{eff} & = & U^\dagger H U \nonumber \\
   & = & H_0 + H_{\sigma\pi\pi}^P 
  + \frac{1}{2}[H_I + H_{\sigma\pi\pi}^P,iS]^P 
              + \frac{1}{3}[[H_I + \frac{1}{2}H_{\sigma\pi\pi}^P,iS],iS]^P + \cdots.
\label{heff}
\end{eqnarray}

\begin{wrapfigure}{r}{4cm}
\centerline{\includegraphics[width=5cm]{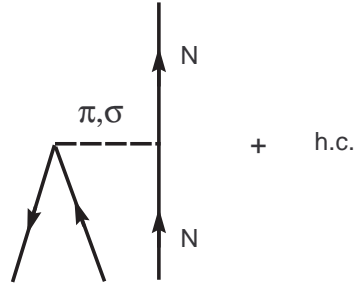}}
\caption{Interaction of "virtual"  process  of order newly generated by the transformation. See the caption of Fig.1.}
\label{fig:2}
\end{wrapfigure}
Here we have omitted the higher order terms in the effective
 Hamiltonian,
 since we restrict ourselves
 to the discussion of the  effective nuclear Hamiltonian in the one-boson-exchange 
approximation. 
The lowest order effective Hamiltonian $H_0$ describes the free energies of
 mesons and nucleons.  It is noticed that we do not find the operators
 for the pion production-absorption by a nucloen, since  the interaction
 Hamiltonian ($H_{\pi NN}$) which produces a virtual pion by a nucleon
 was already eliminated from the Hamiltonian. As a result, elimination
 of the interaction for  'virtual process' generates the effective
 many-body interactions, such as nucleon-nucleon potential,
pion-nucleon potential, the interaction for pion production-absorption
 by two-nucleons, and so on.
 For the purpose of the discussion in the later sections, we shall
 give the explicit form of the
effective interactions in the momentum space.

The third term on the righthand side of Eqn.(\ref{heff}) gives the
 one-boson-exchange nuclear potential $V_{NN}$  expressed as
\begin{eqnarray}
V_{NN} & = & \frac{1}{2}[H_I,iS]^P_{NN}\\
       & = & \frac{1}{2(2\pi)^3}\sum_{M=\pi^b, \sigma} D_{V}
 {}[\bar{u}(\sbm{p}_1')\Gamma_M u(\sbm{p}_1)]
 {}[\bar{u}(\sbm{p}_2') \Gamma_M u(\sbm{p}_2)],
\end{eqnarray}
where the  $\Gamma_M$ stands for the vertex of a meson($M=\pi,\sigma$) and a nucleon ,i.e., 
 $\Gamma_{\pi^b} =  ig \gamma_5 \tau^b$ for $\pi$ and
$ \Gamma_\sigma  =  g$ for $\sigma$, respectively.
The function $D_{V}$ is defined as
\begin{eqnarray}
D_{V} & = & \frac{1}{\Delta_1^2 - \omega_M^2(\sbm{k}_1)}
          + \frac{1}{\Delta_2^2 - \omega_M^2(\sbm{k}_2)}.
\end{eqnarray}
Here we introduced the energy variables as follows: $\Delta_i=E_N(\sbm{p}_i)- E_N(\sbm{p}_{i'})$,
$\sbm{k}_i = \sbm{p}_i - \sbm{p}_{i'}$,
$E_N(\sbm{p})=\sqrt{M_N^2 + \sbm{p}^2}$ and
$\omega_M(\sbm{k})=\sqrt{m_M^2 + \sbm{k}^2}$ ($M=\pi,\sigma$).
It is noticed that $\sbm{p}_i$ and $\sbm{p}_i'$ are momentum operators
operating initial and final nuclear wave functions, respectively.

The remaining terms in Eq. (\ref{heff}) give pion production operator $W$
in one-boson-exchange model, 
\begin{eqnarray}
W &=& 
 \frac{1}{3}[[H_I + \frac{1}{2}H_{\sigma\pi\pi}^P,iS],iS]^P_{NN}
  \\
 &=&  W_\pi + W_\sigma + W_{\pi\sigma},
\end{eqnarray}
which consists of $\pi$-exchange $W_\pi$,
$\sigma$-exchange $W_{\sigma}$ shown in Figs 3(a) and 3(b), 
and $\pi-\sigma$ exchange mechanism  $W_{\pi\sigma}$ shown in Fig. 3(c).
 The explicit form of the production operator of a pion
with momentum $\sbm{q}$ and isospin $a$ in the momentum space is given as
\begin{eqnarray}
W_\pi + W_\sigma & = &
     \frac{1}{3}\frac{1}{(2\pi)^3\sqrt{(2\pi)^32\omega_\pi(\sbm{q})}}
   \sum_{M=\pi^b,\sigma}
   {}[\bar{u}(\sbm{p}_1')
     (\Gamma_{\pi^a}D_{W}\Gamma_M - \Gamma_M D_{W'}\Gamma_{\pi^a})
    u(\sbm{p}_1)]\nonumber \\
 && \times   {}[\bar{u}(\sbm{p}_2') \Gamma_M u(\sbm{p}_2)] + 
( 1 \leftrightarrow 2),
\end{eqnarray}
The interaction for diagrams in Fig. 3(a) is expressed by the term with  $D_W$,
which consists of terms $D_{i+}$ and $D_{i-}$ of the intermediate nucleon states 
with positive  and negative energy, respectively, defined by 
\begin{eqnarray}
 D_W & = & D_{1+} + D_{1-} + D_{2+} + D_{2-}, \nonumber \\
D_{1\pm} & = & 
\Lambda^{\pm}(\sbm{p}_n)\frac{\Delta_{n\pm}-\Delta_2}
        {(\Delta^2_{n\pm}- \omega^2_M(\sbm{k}_2))
         (\Delta^2_2       - \omega^2_M(\sbm{k}_2))}, \nonumber \\
D_{2\pm} & = & 
\Lambda^{\pm}(\sbm{p}_n)
        \frac{1}{\bar{\Delta}_{n\pm}+\omega_\pi(\sbm{q})}
        (\frac{1}{\Delta^2_{n\pm}- \omega^2_M(\sbm{k}_2)} +
         \frac{2}{\Delta^2_2     - \omega^2_M(\sbm{k}_2)}).
\end{eqnarray}
with  $\Lambda^+(\sbm{p})=u(\sbm{p})\bar{u}(\sbm{p})$ and
 $\Lambda^-(\sbm{p})=v(-\sbm{p})\bar{v}(-\sbm{p})$.
The momentum $\sbm{p}_n$ ($ = \sbm{p}_1 + \sbm{k}_2 =
 \sbm{p}_1 + \sbm{p}_2 -\sbm{p}_2'$) denotes the the nucleon momentum in the
   intermediate state. The energy variables   are
 defined as  $\Delta_{n\pm}=E_N(\sbm{p}_1)\mp  E_N(\sbm{p}_n)$, 
and $\bar{\Delta}_{n\pm}=E_N(\sbm{p}_1')\mp E_N(\sbm{p}_n)$.

The interaction for the diagram in Fig. 3(b) is expressed by the term with  $D_{W'}$,
\begin{eqnarray}
 D_W' & = & D_{1+}' + D_{1-}' + D_{2+}' + D_{2-}',
\end{eqnarray}
where the functions $D_{i\pm}'$ are obtained from $D_{i\pm}$
  by replacing $\sbm{p}_n$, $\Delta_{n\pm}$
and $\Delta_{n'\pm}$ by $\sbm{p}_n'=\sbm{p}_1'-\sbm{k}_2$,
$\Delta_{n'\pm}= -E_N(\sbm{p}_1')\pm E_N(\sbm{p}_n')$ and
$\bar{\Delta}_{n'\pm}=-E_N(\sbm{p}_1)\pm E_N(\sbm{p}_n')$, respectively.

\begin{figure}
\centerline{\includegraphics[width=14 cm]{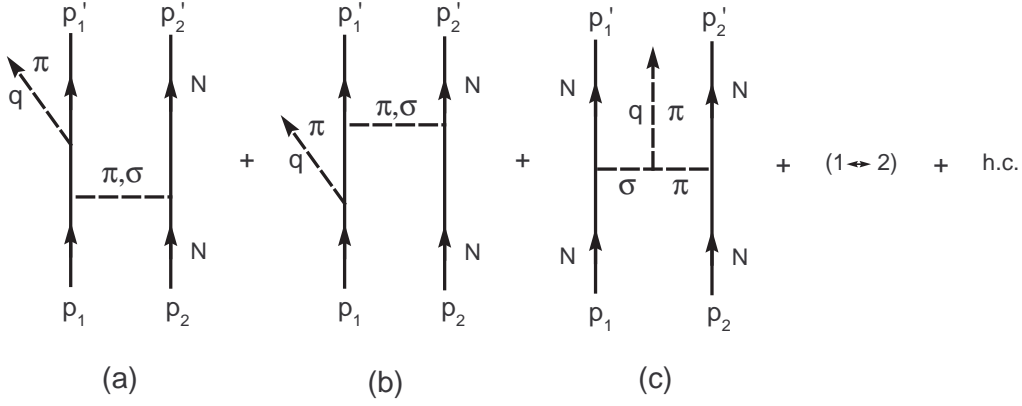}}
\caption{Pion Production by two nucleons.
Figs.(a) and (b) show the nucleon-pole diagrams, while fig.(c) shows
the sigma-pion exchange diagram.}
\label{fig:3}
\end{figure}

Structure of the $D_{1\pm}$ terms is the same as the one of the
 corresponding  diagrams in
 the meson-exchange current\cite{Kobayashi97,Futa76}.
  For the discussion in later section, let us investigate the
 matrix element $D$ at the threshold of pion production ($\sbm{q}=0$) in the center of
 mass system ($\sbm{p}_1=-\sbm{p}_2$ and $\sbm{p}_1'=-\sbm{p_2'}$).  
We notice that $\Delta_{n+}=\Delta_2$ and $D_{1+}$ vanishes. This is
 well known as a fact 
that the recoil and renormalization currents cancel each other for the vanishing 
momentum $\sbm{q}$ in the meson-exchange current. On the other hand the $D_{2+}$ term
 remains finite, and
 can be expressed in terms of the commutator of the "conventional"
 single-nucleon pion production
operator and the nuclear potential.
The pion production operator $(W_\pi + W_\sigma)_+$ with the 
intermediate nucleon state of positive energy at threshold is simply expressed as
\begin{eqnarray}
(W_\pi + W_\sigma)_+ & = & \frac{1}{m_\pi}[H_{\pi NN},V_{NN}]. \label{threw}
\end{eqnarray}
Note that the $D_{1-}$ and $D_{2-}$ terms as well as the $W_{\pi\sigma}$ term remain to be finite.  
Details of the latter term are separately discussed in  Appendix.

As a result, our effective nuclear Hamiltonian
in one-boson-exchange approximation is expressed as follows:
\begin{eqnarray}
H_{eff}=H_0 + H_{\sigma\pi\pi}^P + V_{\pi N} + V_{NN}+W.
\label{eqn:intall}
\end{eqnarray}
It is noticed that we have still the pion degree of freedom as expressed by the terms $V_{\pi N}, W, H^P_{\sigma \pi\pi}$ in our effective Hamiltonian.   The nuclear state vector is determined  by solving the coupled channel equations of the above effective Hamiltonian
in the $NN$ and $\pi NN$ Fock space.

\section{Effective Nuclear Axial-Vector Current}

We shall derive the nuclear axial-vector current,
  by transforming the axial-vector current (\ref{ax_s}) and (\ref{ax_t}) under the {\it same} unitary transformation $U$ as used in deriving the effective nuclear Hamiltonian. This is important to make the theory self-consistent.

\begin{eqnarray}
A_{eff}^{\mu a}(x)& =& U^{\dagger}A^{\mu a}(x) U, \\
 & = & A^{\mu a} + [A^{\mu a},iS]
    +  \frac{1}{2}[[A^{\mu a}+ \frac{1}{3}[ A^{\mu a},iS],iS],iS]\nonumber \\
&&    + [A^{\mu a} + [A^{\mu a},iS],iS'].
\end{eqnarray}

The effective nuclear current consists of  meson current($A^{\mu a}_M$),
the nucleon current ($A^{\mu a}_N$) and meson exchange current
($A^{\mu a}_{NN}$).
Existence of the meson current with the same form as the original one
 is essentially important to describe axial-vector current in the $\pi NN$
Fock space. The nucleon current including pion pole term 
in the momentum space is given as
\begin{eqnarray}
A_N^{\mu a} &=&  
 \bar{u}(\sbm{p}')a^{\mu a}_N(p',p) u(\sbm{p}),
\end{eqnarray}
with
\begin{eqnarray}
a_N^{\mu a}(p',p) &=&  [\gamma^\mu\gamma_5 -
 \frac{2 g f_\pi (p'-p)^\mu }{(p'-p)^2 - m_\pi^2}\gamma_5]\frac{\tau^a}{2}.
\end{eqnarray}
Here four vector $p(p')$ is defined as $p^\mu = (E_N(\sbm{p}),\sbm{p})$.

The meson exchange current consists of pion-exchange current 
$(A^{\mu a}_{NN\pi})$, sigma-exchange current $(A^{\mu a}_{NN\sigma})$ and
pi-sigma-exchange current $(A^{\mu a}_{NN\pi\sigma})$.
The explicit form of the exchange current is given by
\begin{eqnarray}
A^{\mu a}_{NN\pi} +A^{\mu a}_{NN\sigma} 
 & = & \frac{1}{2(2\pi)^3}\sum_{M=\pi^b,\sigma}
 {}[\bar{u}(\sbm{p}_1')(a^{\mu a}_{NN} \Gamma_M - \Gamma_M a^{\mu a}_{NN'})
 u(\sbm{p}_1)]\nonumber \\
{}&&\times  [\bar{u}(\sbm{p}_2') \Gamma_M u(\sbm{p}_1)] + 
( 1 \leftrightarrow 2) ,
\end{eqnarray}
with
\begin{eqnarray}
a^{\mu a}_{NN} & = &
 \bar{a}^{\mu a}_N(p_1',p_{n+})D_{1+} + 
 \bar{a}^{\mu a}_N(p_1',p_{n-})D_{1-} + 
       a^{\mu a}_N(p_1',p_{n-})D_{3-},  \\
a^{\mu a}_{NN'} & = &
 D_{1+}'\bar{a}^{\mu a}_N(p_{n'+},p_1) + 
 D_{1-}'\bar{a}^{\mu a}_N(p_{n'-},p_1) + 
 D_{3-}' a^{\mu a}_N(p_{n'-},p_1),
\end{eqnarray}
and
\begin{eqnarray}
\bar{a}_N^{\mu a}(p',p) &= &[\gamma^\mu \gamma_5
    - \frac{1}{3}\frac{2 g f_\pi}{(p' - p)^2 - m_\pi^2}
      (p' - p)^\mu \gamma_5]  \frac{\tau^a}{2},\\
D_{3-} & = & \Lambda^-(\sbm{p}_n)\frac{1}{\Delta_{n-}+\Delta_2}
  {}[\frac{1}{\Delta^2_{n-}- \omega^2_M(\sbm{k}_2)} +
         \frac{1}{\Delta^2_2     - \omega^2_M(\sbm{k}_2)}].
\end{eqnarray}
Here four-vectors are defined as
 ${p'_1}^{\mu}=(E_N(\sbm{p}_{1}'),\sbm{p}_{1}')$,
$p^\mu_{n\pm} = (\pm E_N(\sbm{p}_n),\sbm{p}_n)$ and
$p^\mu_{n'\pm} = (\pm E_N(\sbm{p}_n'),\sbm{p}_n')$.
It is noticed that the energy variable $E_N(\sbm{p})$ is a function of the momentum operator $\sbm{p}$.
The $D_{1\pm}$ term, which appeared in the pion production operator,
 is exactly the same as a sum of  the recoil, renormalization and pair currents.

Although we have no terms with $D_{2\pm}$, which is  related to
 nuclear potential, in the meson-exchange current, we have a new 
pair-current $D_{3-}$ expressed as a commutator of the 
nucleon-current and the operator
 $S'$ of the unitary transformation associated with the
 interaction shown in Fig.2.
The term $D_{3-}'$ is obtained from $D_{3-}$ by replacing the momenta in the same way  
 as was done for obtaining  $D'_{i\pm}$ in the previous section.
  
We also have the meson-exchange current with $\pi\sigma$ exchange
 mechanism, which is summarized in Appendix, and is discussed
 with the corresponding operator of the pion production. 

Consequently, the total effective nuclear axial-vector current
  consists of the one-nucleon current (the impulse current ) 
 $A_{N}^{\mu a}$,  meson current $A_{M}^{\mu a}$ and the 
 meson-exchange current $A_{NN}^{\mu a}$:
\begin{eqnarray}
A_{eff}^{\mu a}(\sbm{x})=A_{N}^{\mu a}(\sbm{x})+A_{M}^{\mu a}
              (\sbm{x})+A_{NN}^{\mu a}(\sbm{x}).
\end{eqnarray}

\section{Nuclear Axial-Vector Current and PCAC Relation}

The basic Hamiltonian and axial-vector current of the $\sigma$-model satisfies
the PCAC relation.
\begin{eqnarray}
\sbm{\nabla}\cdot \sbm{A}^a + i [H,A^{0 a}] = -f_\pi m_\pi^2 \pi^a.
\end{eqnarray}
In the present treatment, the PCAC relation is transformed as
\begin{eqnarray}
\sbm{ \nabla}\cdot
\sbm{ A}_{eff}^{a}
\ +\ i[\ H_{eff}\ ,A_{eff}^{0 a}\ ]
=-f_{\pi}m_{\pi}^2 \Pi_{eff}^a.
\label{eqn:PCAC}
\end{eqnarray}
The effective pion field $\Pi_{eff}^a$ is given 
by the unitary transformation of the pion field as
\begin{eqnarray}
\Pi_{eff}^a & = & U^\dagger \pi^a U,
\end{eqnarray}
and can be given in one-boson-exchange approximation as
\begin{eqnarray}
\Pi_{eff}^a & = & \pi^a + \Pi^a_N + \Pi^a_{NN}.
\end{eqnarray}
It is easily shown that the effective axial-vector current, the effective
 nuclear Hamiltonian, and the effective pion field satisfy the PCAC relation:
The relation can be divided into three parts,1) meson current, 2) one-nucleon
 current and 3) exchange current:\\

1) The meson current\\  
The meson current ($\sbm{ A}_M^{a},A_M^{0 a}$) satisfies 
\begin{eqnarray}
\sbm{ \nabla}\cdot \sbm{ A}_M^{a}+i[ H_0,A_M^{0 a} ]=-f_{\pi}m_{\pi}^2\pi^a.
\end{eqnarray}
It is noticed this is a consequence of existence of the meson degrees of freedom in our nuclear Hamiltonian. Without these freedoms, one cannot satisfy the PCAC relation.\\
2) The one-nucleon current\\
The one-body current satisfies the relation very familiar to us:
\begin{eqnarray}
\sbm{ \nabla}\cdot
\sbm{ A}_N^{a}+i[ H_0 , A_N^{0 a} ] =-f_{\pi}m_{\pi}^2\Pi_N^a,
\end{eqnarray}
with
\begin{eqnarray}
\Pi_N^a & = & \frac{1}{(p'-p)^2 - m_\pi^2}\bar{u}(\sbm{p}')\Gamma_{\pi^a}
u(\sbm{p}),
\end{eqnarray}
3) The meson-exchange current\\
 The PCAC relation for the meson-exchange current is non-trivial.
 It is expressed as follows:
\begin{eqnarray}
\sbm{ \nabla}\cdot
\sbm{ A}_{NN}^{a}
  + i\Bigl[ H_0 , A_{NN}^{0 a}\Bigr]
  + i\Bigl[ V_{NN} , A_{N}^{0 a}\Bigr]
  + i\Bigl[ W_{\pi} , A_M^{0 a}\Bigr]
\nonumber \\
 = -f_{\pi}m_{\pi}^2 \Pi_{NN}^a.
\end{eqnarray}
Here it is noticed that the above equation includes the meson-exchange current, the
one-nucleon current and also meson-current. Thus, in order to complete the PCAC relation,
we need the meson current and the pion production-absorption operator
 in addition to the nuclear potential and meson-exchange current.
This fact has been overlooked in the previous works so far done.

\section{Relation of Pion Production and the PCAC}

We shall investigate the relation of pion production reaction and the
 PCAC
 relation. In order to
avoid unnecessary complications, we shall assume the nuclear
 system with two nucleons and pions.
The relation of the one-body axial-vector current in Eqs. (4.6) and (4.7) 
suggests that one can obtain pion production operator by using the PCAC relation.
However the PCAC relation connects  divergence of the axial-vector current 
with the the effective pion field operator, but not the pion production operator to be used in the dynamical equations describing the nuclear system with nucleons and pions.
By using the reduction formula, we have a relation  between the matrix element
 of the divergence of the
axial-vector current and the pion production T-matrix element as
\begin{eqnarray}
i q^\mu \langle NN|A_\mu^a|NN\rangle & = & \frac{f_\pi m_\pi^2}{q^2-m_\pi^2}
  \sqrt{(2\pi)^3 2\omega_\pi(k)}T_{NN \rightarrow \pi NN}
\end{eqnarray}
with  the pion momentum  $q^\mu$, which should hold for
 any formalism to describe nuclear system. It is noticed that 
the $NN\rightarrow \pi NN$ T-matrix is calculated by solving the coupled 
channel equations for the nuclear system. 
At first sight, one may point out that this equality contradicts with our present theory,
because the left-hand side of the equation includes the matrix elements of the one-nucleon
operator, while the right-hand side of the equation does not.
Absence of the one-nucleon operator for the pion production, however,  does not lead to any
contradiction, since the pion in the final state 
is always produced from  a nucleon interacting with other nucleons. In other words, this observation shows that we evaluate the matrix element of the two-nucleon operator such as a product
 of the one-boson-exchange potential and the one-nucleon pion production operator,  
 which corresponds to our pion production operator $(W_\sigma + W_\pi)$ . 
 
To show  that this is really the case,
we shall investigate the matrix element of the pion production operator
 $(W_\sigma + W_\pi)_+$ at threshold. (Since other terms 
in $(W_\sigma + W_\pi)$ are essentially irrelevant to the 
two-nucleon potential, we shall safely skip them from the discussion 
in what follows.) 
Now we calculate  the T-matrix element by using the
solution of the coupled channel equation in the lowest order:
\begin{eqnarray}
T_{NN \rightarrow \pi NN} & \sim & <NN_f(DW)|(W_\pi + W_\sigma)_+|NN_i(DW)>. \label{piNN}
\end{eqnarray}
Here the $|NN_{i/f}(DW)>$ are the nuclear state vectors with
energy $E_{i/f}$ obtained by solving Schr\"{o}dinger equation.
Using Eq. (\ref{threw}) and noticing $[H_0, H_{\pi NN}]=0$
 at threshold($E_i - E_f=m_\pi$),
we can rewrite Eq. (\ref{piNN}) as
\begin{eqnarray}
T_{NN \rightarrow \pi NN} & = &
     <NN_f(DW)|\frac{1}{m_\pi}[H_{\pi NN},H_0+ V_{NN}]|NN_i(DW)>\\
   & = &  <NN_f(DW)|H_{\pi NN}|NN_i(DW)>.
\end{eqnarray}

This is just the matrix element of the pion source function of the
 one-nucleon type.
 Therefore, our pion-production operator is consistent with the PCAC relation.

The present result also implies that the PCAC relation requires us to construct  
the nuclear force, pion production operator and exchange currents in a
consistent way,
 and also to describe  the nuclear system including the pion degree of
 freedom
 based on the nuclear force thus derived.

\section{Summary}

We have studied the nuclear PCAC relation in the recent theory of the effective 
nuclear interactions which include meson degrees of freedom explicitly.
First of all, in order to make the discussion clear, we assumed the
 linear $\sigma$-model, and derived the nuclear force, the effective
 interaction
 for the  pion production-absorption and
the nuclear axial-vector current by applying the unitary transformation.
 All the operators we have derived are Hermitian and are functions of
 the local momenta of the relevant particles. 
Secondly, we have studied in detail the nuclear PCAC relation , in
particular,
 relationship among  the pion-production mechanism, the axial-vector
 current
 and the nuclear force.
As results, 1) it was shown that one has to take into account the
explicit
 pion and sigma meson degrees of freedom in addition to the nucleons
 in order to satisfy the PCAC relations. In particular, a consistent
 description of the nculear force,
and also pion production operator played an essential role to hold the PCAC relation.
2) The PCAC relation gives us just the effective pion field, but
  not the  interactions for the pion production-absorption processes.
3) The fact that the one-body pion production operator does not
 exist in our formalism means simply that a free nucleon cannot emit
 or absorb a pion. As a consequence, the effective interaction
 operator for pion production-absorption  includes the term
 which corresponds to production of a pion by a nucleon
 interacting  with other nucleons, in addition to the conventional
 irreducible two-nucleon operators.
This looks similar to the traditional approach which incorporate
 the one-nucleon operator plus  nuclear correlations, as far as we concern ourselves with  the pion production at threshold.

It is noticed that the above observations about  the nuclear PCAC
 relation holds  independent of any strong interaction models
 which lead the PCAC relation. 
Equations (4.8) and (5.1)  showed that the PCAC relation for
 the two-nucleon axial-vector current is not a simple
 relation assumed by Blin-Stoyle and Tint for the Gamow-Teller operator. 

  To make the nuclear axial-vector current and pion production-absorption
  operators more realistic,  we should incorporate  short-range
  mechanism into the present work starting, for example
  from the  model based on chiral symmetry\cite{Adam,Ana02}.

\section*{Acknowledgements}

This work was supported  by the 
Japan Society for the Promotion of Science, 
Grant-in-Aid for Scientific Research (C) 15540275.

\appendix
\section{$\sigma\pi$ exchange current and pion production operator}

Here we give  full expressions for the $\sigma-\pi$ exchange current
and pion production operator.
The pion production operator is given as
\begin{eqnarray}
W_{\pi\sigma} & = &
\frac{m_\sigma^2 - m_\pi^2}{f_\pi (2\pi)^3\sqrt{(2\pi)^3 2\omega_\pi(\sbm{q})}}
 {}[\bar{u}(\sbm{p}_1')\Gamma_\sigma u(\sbm{p}_1)]
 {}[\bar{u}(\sbm{p}_2')\Gamma_{\pi^a} u(\sbm{p}_2)]D_{\pi\sigma}
 + ( 1 \leftrightarrow 2 ), \nonumber \\
\end{eqnarray}
where 
\begin{eqnarray}
D_{\pi\sigma} & = & D_M [ 1 +
 \frac{\Delta_1 + \Delta_2 - \omega_\pi(\sbm{q})}
      {12 \omega_\sigma \omega_\pi}( 
 \frac {(\Delta_1 -\omega_\sigma)(\Delta_2 -\omega_\pi)}
      {\omega_\sigma + \omega_\pi + \omega_\pi(\sbm{q})}
\nonumber \\
&&
- \frac{(\Delta_1 +\omega_\sigma)(\Delta_2 +\omega_\pi)}
      {\omega_\sigma + \omega_\pi - \omega_\pi(\sbm{q})}
- \frac{(\Delta_1 -\omega_\sigma)(\Delta_2 +\omega_\pi)}
      {\omega_\sigma - \omega_\pi + \omega_\pi(\sbm{q})}
    )]
\end{eqnarray}
with
\begin{eqnarray}
D_M & = & \frac{1}{(\Delta_1^2 - \omega_\sigma^2)(\Delta_2^2 - \omega_\pi^2)}
\end{eqnarray}
Here we denote $\omega_\sigma=\omega_\sigma(\sbm{k}_1)$ and
$\omega_\pi=\omega_\pi(\sbm{k}_2)$.
The second term in the square bracket proportional to 
$\Delta_1 + \Delta_2 - \omega_\pi(\sbm{q})$ vanishes in the Born
approximation because of the energy conservation, and it contributes to the
off -energy shell matrix element.

The axial-vector exchange current is given as
\begin{eqnarray}
A_{NN\pi\sigma}^{\mu a} & = &
\frac{1}{(2\pi)^3} D_M^{\mu}
 {}[\bar{u}(\sbm{p}_1')\Gamma_\sigma u(\sbm{p}_1)]
 {}[\bar{u}(\sbm{p}_2')\Gamma_{\pi^a} u(\sbm{p}_2)]
 + ( 1 \leftrightarrow 2 ).
\end{eqnarray}
The non pion-pole term of the exchange current is simply given as
\begin{eqnarray}
D_{M}^{\mu } & = & i (p_1 - p_1' - p_2 + p_2')^\mu D_M,
\end{eqnarray}
while the  pion-pole term is given as
\begin{eqnarray}
\sbm{D}_{M} & = &
-i (\sbm{p}_1 + \sbm{p}_2 - \sbm{p}_1' - \sbm{p}_2') \nonumber \\
&& \times \frac{m_\sigma^2 - m_\pi^2}{6\omega_\sigma\omega_\pi} 
(\frac{\Delta_1 \Delta_2 + \omega_\sigma\omega_\pi}
{(\omega_\sigma+\omega_\pi)^2 - \omega_\pi(\sbm{q})}
 +
\frac{\Delta_1 \Delta_2 - \omega_\sigma\omega_\pi}
{2\omega_\pi(\sbm{q})(\omega_\sigma-\omega_\pi + \omega_\pi(\sbm{q}))})D_M,
 \\
D^0_{M} & = &
-i \frac{m_\sigma^2 - m_\pi^2}{6\omega_\sigma\omega_\pi}
%\nonumber \\
%&&
(\frac{(\omega_\sigma+\omega_\pi)
  (\Delta_1 \omega_\pi + \Delta_2 \omega_\sigma)}
{(\omega_\sigma+\omega_\pi)^2 - \omega_\pi(\sbm{q})^2}
 -
\frac{\Delta_2\omega_\sigma - \Delta_1\omega_\pi}
{2(\omega_\sigma-\omega_\pi + \omega_\pi(\sbm{q}))})D_M.
 \nonumber \\
\end{eqnarray}

\end{document}